\documentclass[%
 reprint,
 amsmath,amssymb,
 aps,
]{revtex4-1}

\usepackage{graphicx}% Include figure files
\usepackage{dcolumn}% Align table columns on decimal point
\usepackage{bm}% bold math
\usepackage{hyperref}% add hypertext capabilities
\begin{document}

\preprint{APS/123-QED}

\title{Suppression of Overfitting in Extraction of Spectral Data from Imaginary Frequency Green Function Using Maximum Entropy Method}% Force line breaks with \\

\author{Enzhi Li}
\affiliation{Suning R\&D Center, Palo Alto, USA}
 \altaffiliation[Also at ]{Department of Physics \& Astronomy, LSU, USA}
  \email{enzhililsu@gmail.com}

\date{\today}

\begin{abstract}
Although maximum entropy method (maxEnt method) is currently the standard algorithm for extracting real frequency information from imaginary frequency Green function, yet this method is beset with overfitting problem, which manifests itself as the spurious spikes in the resultant spectral functions. To address this issue and motivated by the regularization techniques widely used in machine learning and statistics, here we propose to add one more regularization term into the original maxEnt loss function to suppress these redundant spikes. The essence of this extra regularization term is to demand that the resultant spectral functions should pay a price for being spiky. We test our algorithm with both artificial and real data, and find that spurious spikes in the resultant spectral functions can be effectively suppressed by this method. 
\end{abstract}

\maketitle

%\tableofcontents

\section{Introduction}
The extraction of spectral data from imaginary frequency Green function, or the analytic continuation of Green function from imaginary frequency space to real frequency space, is a central problem in Monte Carlo simulation in that only through the real frequency spectral function can we make a comparison between Monte Carlo simulations and experimental results. Several methods have been proposed to achieve this purpose, such as Pad\'e approximation\cite{gunnarsson2010analytical, beach2000reliable}, least square fit\cite{schuttler1985monte, schuttler1986monte}, stochastic analytic continuation\cite{sandvik1998stochastic, sandvik2016constrained}, and maximum entropy method\cite{gull1988bayesian, silver1990maximum, jarrell1996bayesian}. In general, Pad\'e approximation cannot yield numerically stable spectral data, and is thus rarely used now. Least square method which can be viewed as an attempt to find the spectral function via direct inversion of a kernel matrix (to be defined later) has achieved limited success due to the ill-conditioned nature of the kernel matrix, and is now replaced by the maximum entropy (maxEnt) method. Currently, maxEnt method is the paradigmatic algorithm for implementing analytic continuation, although people have shown that maxEnt can be considered as a special case of stochastic analytic continuation method\cite{beach2004identifying}. 

MaxEnt method can be roughly viewed as a least square method with a regularization term\cite{fuchs2010analytic}. Similar to least square method, maxEnt aims to find a curve that best fits the data at hand. What distinguishes maxEnt from least square method is that maxEnt picks among all possible candidate spectral curves the best one that could maximize the entropy (regularization) term. The principle of maximum entropy is a mathematical formulation of our intuition that we do not arrogate any extra knowledge other than the information that we have at hand\cite{jaynes1957information}. Combination of the least square method and the principle of maximum entropy yields the maxEnt algorithm which gives us the optimal inference on spectral data given incomplete information about imaginary frequency Green function. 

MaxEnt method, successful although it is, is still beset with the overfitting problem which manifests itself in the form of spurious peaks that appear in the resultant spectral curves\cite{otsuki2017sparse}. In the field of statistics and machine learning, overfitting is obnoxious and ubiquitous, and  people have devised various methods to solve this problem, such as adding regularization terms in curve fitting, dropping with some probability the hidden units in deep learning, etc\cite{friedman2001elements, goodfellow2016deep, hinton2012improving, srivastava2014dropout}. Motivated by the regularization methods that are widely use in statistics and machine learning, here we propose a method to reduce or even to eliminate overfitting in maxEnt. In order to suppress overfitting, we propose to introduce an extra regularization term into the original maxEnt method to penalize the spectral curves on their spikiness. The smoothness condition was previously imposed on spectral curves that are generated using least square method\cite{white1989monte, jarrell1989dynamical, sandvik1998stochastic}, and here we further propose to impose this condition on maxEnt method. We make comparisons between the classical maxEnt method and the maxEnt method with this extra regularization term, and show definitely that our regularization term can effectively reduce or even to eliminate the spurious peaks we often see in classical maxEnt results. 

The organization of the paper is as follows. In section II, we will summarize the classical maxEnt method. Our goal in this section is to show that this method is essentially a least square method with a regularization (entropy) term, and can be converted to a convex optimization problem, for which standard numerical recipes are readily available\cite{boyd2004convex}. The algorithm described in this section is implemented using Python. In section III, we will extend the classical maxEnt by introducing another regularization term to solve the overfitting problem. The advantage of this extended version of maxEnt is that it not only suppresses the spurious peaks in the resultant spectral curves, but also preserves the convexity of the problem. Moreover, only slight modifications of the maxEnt program are needed to implement the new algorithm as described in this section. We also implement a Python program for this section. Comparisons with classical maxEnt method are made here. In section IV, we test our algorithm with Monte Carlo simulation data of symmetric periodic Anderson model, the spectral results of which are abundant and well known, and can be used to benchmark our method. Programmatic results show that we can accurately reproduce the three-peak structure for the localized electron spectral data, and spurious spikes that arise in the classical maxEnt method are effectively eliminated by the extra regularization term we introduce here. A conclusion is made in section V. Detailed mathematical derivations and further discussion of our method can be found in the appendix of this paper. 

\section{Classical maximum entropy method}
\label{classicalMaxEnt}
Through quantum Monte Carlo simulation, we can obtain Green function in imaginary time or imaginary frequency space. Imaginary frequency Green function is related to the imaginary time Green function through a Fourier transformation: 
\begin{eqnarray}
G(i \omega_n) = \int_{0}^{\beta}G(\tau) e^{i\omega_n\tau}d\tau, 
\end{eqnarray}
where $\beta = \frac{1}{T}$ is the inverse temperature ($k_B = 1$), $\omega_n = \frac{(2n+1)\pi}{\beta}$ for Fermions and $\omega_n = \frac{2n\pi}{\beta}$ for Bosons. The spectral representation of Green function is 
\begin{eqnarray}
\label{spectral_representation}
G(i\omega_n ) = \int_{-\infty}^{\infty}\frac{A(\omega)}{i\omega_n - \omega} d\omega, 
\end{eqnarray}
where $A(\omega)$ is the spectral function and satisfies the normalization condition $\int_{-\infty}^{\infty} A(\omega) d\omega = 1$. Once we know the spectral function, we can obtain the time-retarded real frequency Green function through the replacement $G^{+}(\omega) = G(i\omega_n \rightarrow \omega + i 0^{+})$. The time-retarded Green function is directly related with experimental measurements and is thus the final target of Monte Carlo simulations.The naive replacement of $i\omega_n$ with $\omega + i 0^{+}$ is not viable in practice since we can only obtain the numerical values, rather than the analytic form, of imaginary frequency Green function from Monte Carlo simulation. Worse still, the numerical results are inevitably compromised to some degree by all kinds of noises, which renders the direct inversion of Equation [\ref{spectral_representation}] impossible. Maximum entropy method is designed to extract the spectral function from Equation [\ref{spectral_representation}] without resort to direct inversion. Just as least square method can be interpreted using Gaussian distribution, maxEnt can also be interpreted in the language of Bayesian inference\cite{jarrell1996bayesian}. Here, we prefer to view the maxEnt as a generalization of least square method. The essence of maxEnt is to choose among all possible spectral functions the one that best fits the measured data. Assume that we already have the spectral function $A(\omega)$, then the corresponding Green function could be evaluated as 
\begin{eqnarray}
G(i\omega_n) &=& \int_{-\infty}^{\infty}\frac{A(\omega)}{i\omega_n - \omega}d\omega \\\nonumber
& := & KA 
\end{eqnarray}
During Monte Carlo simulation process, to reduce the correlation between successive measurements, we perform $N_s$ Monte Carlo updates before making a measurement. The number $N_s$ is chosen in such a way that the cross-correlation between successive measurements is negligible. Therefore, the number of measurements is equal to the total number of Monte Carlo updates divided by $N_s$. This splitting strategy of Monte Carlo steps is called binning, and the number of actual measurements is denoted  as the number of bins $N_b$. The final measured Green function is the bin average of these measured Green functions: 
\begin{eqnarray}
\bar{G}(i\omega_n) = \frac{1}{N_{b}}\sum_{i = 1}^{N_b}G^{(i)}(i\omega_n)
\end{eqnarray}
The covariance matrix of these binned Green functions is
\begin{eqnarray}
C_{mn} &=& \frac{1}{N_b(N_b - 1)}\sum_{i=1}^{N_b} \Big(G^{(i)}(i\omega_m) - \bar{G}(i\omega_m)\Big)^{*}\\ \nonumber
&\times& \Big(G^{(i)}(i\omega_n) - \bar{G}(i\omega_n)\Big)
\end{eqnarray}
It is obvious from the above definition that the covariance matrix is Hermitian and (maybe semi) positive definite. Since we have already carefully chosen the number of $N_s$ to make sure the cross-correlation between successive measurements is negligible, we can assume that the covariance matrix is real. This assumption can significantly simplify the derivation of maxEnt algorithm. In the appendix, there is a detailed discussion about the influence of the imaginary part of covariance matrix on our algorithm. 

Given the spectral function $A(\omega)$ and bin averaged Green function $\bar{G}(i\omega_n)$, we can define the squared error as 
\begin{widetext}
\begin{eqnarray}
\chi^2 = \sum_{mn}\Bigg( \bar{G}(i\omega_m) - \int_{-\infty}^{\infty}\frac{A(\omega)}{i\omega_m - \omega}d\omega \Bigg)^{\dagger} C^{-1}_{mn}  \Bigg(\bar{G}(i\omega_n) - \int_{-\infty}^{\infty}\frac{A(\omega)}{i\omega_n - \omega}d\omega \Bigg)
\end{eqnarray}
\end{widetext}
If we only try to minimize the squared error, then our algorithm is nothing but the primitive least square method. In reality, this naive algorithm cannot yield numerically stable results for our data. MaxEnt adds to this squared error another term that tries to regularize the final results. If we think of $A(\omega)$ as describing some probability distribution, then this regularization term is proportional to the Shannon information entropy of $A(\omega)$ relative to a default model $D(\omega)$: 
\begin{eqnarray}
S = -\int_{-\infty}^{\infty} A(\omega) \log\frac{A(\omega)}{D(\omega)} d\omega, 
\end{eqnarray}
where $D(\omega)$ is the default model which is generally set to be a featureless Gaussian distribution unless we have some \textit{a priori} knowledge about the features of $A(\omega)$. The principle of maximum entropy as proposed in Ref. [\cite{jaynes1957information}] states that if we have no other information about $A(\omega)$, then we should choose the spectral function that can maximize the information entropy $S$. Combination of the least square method and the principle of maximum entropy gives us a loss function which is 
\begin{eqnarray}
Q = \frac{1}{2}\chi^2 - \alpha S
\end{eqnarray}
Here, $\alpha$ is a real parameter that tunes the competition between the tendency to minimize squared error $\chi^2$ and the tendency to maximize entropy $S$. We can understand the squared error $\chi^2$ as system energy and $\alpha$ as temperature. Thus, $Q$ can be interpreted as free energy, and minimization of loss function is equivalent to minimization of free energy in thermodynamics. 

In order to simplify notation, we denote $\xi = \bar{G} - KA$. Since $\xi$ is a complex vector, we can separate its real from its imaginary part as 
\begin{eqnarray}
\xi &=& \xi_{R} + i\xi_{I} \\\nonumber
&=& \bar{G}_{R} - K_{R} A + i(\bar{G}_{I} - K_{I} A)
\end{eqnarray}
Here, we have already employed the notation that 
\begin{eqnarray}
&& K_{R} A = -\int_{-\infty}^{\infty} \frac{\omega A(\omega)}{\omega_n^2 + \omega^2}d\omega \\\nonumber
&& K_{I} A = -\int_{-\infty}^{\infty} \frac{\omega_n A(\omega)}{\omega_n^2 + \omega^2}d\omega
\end{eqnarray}
Similarly, we can also separate the real part and imaginary part of the inverse of covariance matrix as $C^{-1} = C^{-1}_{R} + i C_{I}^{-1}$. Since the inverse of covariance matrix is also Hermitian, $C_{R}^{-1}$ is a real symmetric matrix and $C_{I}^{-1}$ is a real anti-symmetric matrix. With the introduction of these notations, $\chi^2$ can then be recast into this form:
\begin{eqnarray}
\chi^2  = \left( \begin{array}{cc}  \xi_{R}^{T} & \xi_{I}^{T} \\ \end{array} \right) \left( \begin{array}{cc}  C_{R}^{-1} & -C_{I}^{-1} \\  C_{I}^{-1} & C_{R}^{-1} \\ \end{array} \right) \left( \begin{array}{c}  \xi_{R} \\  \xi_{I} \\ \end{array} \right) 
\end{eqnarray}

As noted above, if the correlation between different bins of Green functions is negligible, then we can ignore the imaginary part of covariance matrix, and $\chi^2$ can be simplified as 
\begin{eqnarray}
\chi^2 = \left( \begin{array}{cc}  \xi_{R}^{T} & \xi_{I}^{T} \\ \end{array} \right) \left( \begin{array}{cc}  C_{R}^{-1} & 0 \\  0 & C_{R}^{-1} \\ \end{array} \right) \left( \begin{array}{c}  \xi_{R} \\  \xi_{I} \\ \end{array} \right) 
\end{eqnarray}

Up to now, we have made the implicit assumption that the covariance matrix is invertible. However, in reality, covariance matrix $C$ may contain exceedingly small eigenvalues that are below machine precision. In this case, the calculation of $C_{R}^{-1}$ may be numerically unstable. In order to avoid this, we are to diagonalize the matrix $C_{R}$ and then discard the eigenvalues that are below machine precision. Since $C_{R}$ is real and symmetric, we can always find an orthogonal matrix $U$ such that $C_{R} = U \Lambda U^{T}$, where $\Lambda$ is diagonal. Some of the diagonal elements of $\Lambda$ may be zero (below machine precision), and thus $\Lambda^{-1}$ may diverge. To eliminate this divergence, we shall truncate the matrix $\Lambda$ such that all the diagonal elements are non-zero (above machine precision). The truncated matrix which we denote as $\tilde{\Lambda}$ is now invertible. Similarly, we should also replace $U$ with its truncated counterpart $\tilde{U}$. Now we can replace the possibly singular matrix $C_{R}$ with the invertible (truncated) matrix $\tilde{C_{R}} := \tilde{U}\tilde{\Lambda}\tilde{U}^T$. We should also truncate the vectors $\xi_{R}, \xi_{I}$ to match the dimension of the truncated covariance matrix. Since the singularity of covariance matrix can be resolved by matrix truncation, from now on, we will assume that the covariance matrix is always invertible. 

By introducing an orthogonal matrix $U$ such that $C_{R} = U\Lambda U^{T}$, where $\Lambda$ is diagonal, we can rewrite $\chi^2$ as 
\begin{eqnarray}
\chi^2 = \tilde{\xi}^{T} \left(
\begin{array}{cc}
 \Lambda^{-1} & 0 \\
 0 &  \Lambda^{-1} \\
\end{array}
\right)\tilde{\xi} 
\end{eqnarray}
Here, we have defined 
\begin{eqnarray}
\tilde{\xi}_{R} &=& U^T\xi_R \\\nonumber
&=& U^T G_{R} - U^T K_{R} A \\\nonumber
&=& \tilde{G}_{R} - \tilde{K}_{R}A , \\\nonumber
\tilde{\xi}_{I} &=& U^T\xi_I \\\nonumber
&=& U^T G_{I} - U^T K_{I} A \\\nonumber
&=& \tilde{G}_{I} - \tilde{K}_{I}A , \\\nonumber
\tilde{\xi} &=& \left(
\begin{array}{c}
 \tilde{\xi}_{R} \\
 \tilde{\xi}_{I} \\
\end{array}
\right) 
\end{eqnarray}

With these notations, loss function can be rewritten as  
\begin{eqnarray}
\label{objective_function}
Q &=& \frac{1}{2} \chi^2 - \alpha S \\\nonumber
&=&  \frac{1}{2} \tilde{\xi}_{R}^T\Lambda^{-1}\tilde{\xi}_{R} + \frac{1}{2}\tilde{\xi}_I^T\Lambda^{-1}\tilde{\xi}_{I} - \alpha S
\end{eqnarray}
Now we have obtained a loss function which is actually a functional with $A(\omega)$ as its variable. The functional dependence of $Q$ on $A(\omega)$ is hidden in $\tilde{\xi}_{R}, \tilde{\xi}_{I}$ and $S$. Our aim is to find a spectral function that can minimize this loss function. To accomplish this, we should set the gradient of $Q$ with respect to $A(\omega)$ to be zero, that is, $\nabla_{A(\omega)}Q = 0$, and then find the solution to this equation. Solution of this equation requires numerical computation. In practice, we need to discretize the continuous variable $\omega$ before we can perform any numerical computation. After discretization, the continuous real frequency $\omega$ adopts an integer index. To distinguish the discretized real frequency from the imaginary frequency which already has an integer index, we will use the convention that Greek letters such as $\mu, \nu$ are used to index real frequencies, whereas Latin letters are used to index imaginary frequencies. With discretized real frequencies, the gradient of loss function with respect to spectral function turns into a finite-dimensional vector, which is  
\begin{eqnarray}
f(A(\omega_{\mu})) &=& \frac{\delta Q}{\delta A(\omega_{\mu})} \\\nonumber
&=& \alpha \Big(1 + \log\frac{A(\omega_{\mu})}{D(\omega_{\mu})}\Big) \Delta\omega \\\nonumber
&& - \sum_{nm} \tilde{K}_{R}(\omega_n, \omega_{\mu})\Delta\omega (\Lambda^{-1})_{nm}\Big(\tilde{G}_{R} - \tilde{K}_{R}A\Big)_m \\\nonumber
&& - \sum_{nm} \tilde{K}_{I}(\omega_n, \omega_{\mu}) \Delta\omega (\Lambda^{-1})_{nm}\Big(\tilde{G}_{I} - \tilde{K}_{I}A \Big)_{m}
\end{eqnarray}
Next we are to solve the equation $f(A(\omega_{\mu})) = 0$ to find the optimal spectral function. We can employ Newton's iteration method to solve this equation numerically, that is, we want to solve this equation iteratively: 
\begin{eqnarray}
A_{n+1}(\omega_{\mu}) = A_{n}(\omega_{\mu}) - \sum_{\nu}(H^{-1})_{\mu\nu}f(A_{n}(\omega_{\mu}))
\label{Newton_iteration}
\end{eqnarray}
Here, $H$ is the Hessian matrix of $Q$ with respect to spectral function, which is
\begin{eqnarray}
H_{\mu\nu} &=& \frac{\delta f(A(\omega_\mu))}{\delta A(\omega_\nu)} \\\nonumber
&=& \frac{\delta^2Q}{\delta A(\omega_{\mu})\delta A(\omega_{\nu})} \\\nonumber
&=& \alpha\Delta\omega\frac{\delta_{\mu\nu}}{A(\omega_\mu)} \\\nonumber
&& +  \sum_{nm}\tilde{K}_{R}(\omega_n, \omega_{\mu}) \Delta\omega(\Lambda^{-1})_{nm}\tilde{K}_{R}(\omega_m, \omega_{\nu})\Delta\omega \\\nonumber
&& + \sum_{nm}\tilde{K}_{I}(\omega_n, \omega_{\mu}) \Delta\omega(\Lambda^{-1})_{nm}\tilde{K}_{I}(\omega_m, \omega_{\nu})\Delta\omega 
\end{eqnarray}
It is easy to see that Hessian matrix is positive definite. Thus, the loss function $Q$ is convex and we can always find a global optimal solution to the loss function. In reality, we find that the direct inversion of Hessian matrix may yield unstable results. Moreover, it is much more difficult to parallelize matrix inversion than to parallelize matrix multiplication. As a result of these two considerations, we choose to employ conjugate gradient algorithm\cite{hestenes1952methods} to numerically solve Equ. [\ref{Newton_iteration}]. For the conjugate gradient method to be applicable, we shall rewrite Equ. [\ref{Newton_iteration}] as 
\begin{eqnarray}
\sum_{\nu} H_{\mu\nu} \Big( A_{n+1}(\omega_{\nu}) - A_{n}(\omega_{\nu}) \Big) = -f(A_{n}(\omega_{\mu}))
\end{eqnarray}
The above equation can be understood as a a system of simultaneous linear equations, with the coefficient matrix being positive definite, and thus we can use conjugate gradient method to find the optimal spectral function. 

When running the program to find an $A(\omega)$ that best fits the known data, we start from a very large $\alpha$ value and decrease $\alpha$ exponentially to a tiny value. This technique is called annealing. As noted above, we can interpret the $\alpha$ as temperature. Minimization of loss function $Q$ or free energy is easier at high temperature thanks to the fact that when temperature is high enough, minimization of free energy is essentially equivalent to maximization of entropy. It can be seen from the definition of $Q$ that when $\alpha\rightarrow\infty$, the extremal values of $Q$ coincide with the extremal values of $S$. Setting the gradient of entropy to zero, we have 
\begin{eqnarray}
\frac{\delta S}{\delta A(\omega)} = -1 - \log \frac{A(\omega)}{D(\omega)} = 0
\end{eqnarray}
Solving the above equation yields $A(\omega) = e^{-1}D(\omega)$. Therefore, when $\alpha$ is very large, the optimal spectral function should be almost identical to the default model after normalization condition is imposed. With the spectral result at large $\alpha$, we can use $A(\omega)$ for large $\alpha$ to initialize Newton's iteration method for small $\alpha$ and repeat this procedure until $\alpha$ is small enough and the resultant spectral function no longer changes substantially. This is a vague method of selecting $\alpha$. The selection of $\alpha$ in maxEnt is a long standing problem and three algorithms have been proposed for this purpose, which are the historic, the classic and Bryan's method\cite{jarrell1996bayesian}. Although these methods, especially the classic and Bryan's method, are brilliantly successful, they depend on an arbitrary choice of $p(\alpha|\bar{G})$, which is the probability distribution of $\alpha$ given $\bar{G}$. The elimination of this dependence is the topic of next section.  

We have developed a Python program to implement this algorithm. The program can be found here: \url{https://github.com/PrimerLi/maxEntLambda}. We have tested the program with lots of data, both real and artificial, and found data of higher quality (smaller noise or correlation) yields better spectral functions, just as we anticipated. 

\begin{figure}[h!]
\centerline{\includegraphics[scale = 0.35]{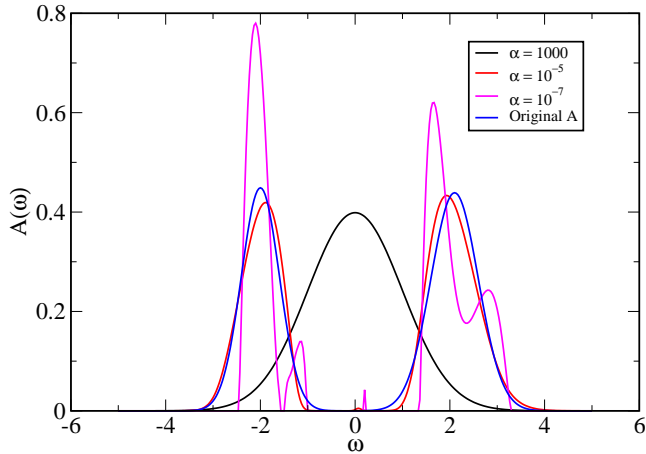}}
\caption{Comparison of the calculated spectral functions for different values of $\alpha$ with the original known spectral function. We have used a featureless Gaussian as our default model. When $\alpha = 1000$, the resultant $A(\omega)$ is indistinguishable from the default model. As $\alpha$ decreases, the resultant spectral functions evolve toward the original spectral function, as anticipated. When $\alpha = 10^{-7}$, overfitting emerges. }\label{A_alpha}
\end{figure}

Here, we present an example spectral function for artificial data. We start with a spectral function $A(\omega) = 0.45 N(\omega, -2,. 0.4) + 0.55 N(\omega, 2.1, 0.5)$, where, $N(\omega, \mu, \sigma) = \frac{1}{\sqrt{2\pi}\sigma} e^{-\frac{(\omega - \mu)^2}{2\sigma^2}}$. From this spectral function, we generate an imaginary frequency Green function, to which we add Gaussian noise with $\mu = 0, \sigma = 0.001$. We then use our program to find an optimal spectral function that best fits the Green function. We start with $\alpha = 1000$ and decrease its value exponentially to $\alpha_f  = 10^{-8}$. Snapshots of these spectral functions for different values of $\alpha$ are shown in Fig. \ref{A_alpha}. We have used standard normal distribution as the default model. From the figure, we can see that when $\alpha$ is very large, the resultant spectral function is indistinguishable from the default model. As $\alpha$ decreases, $A(\omega)$ evolves toward the original spectral function, thus validating our program. When $\alpha$ is too small, overfitting emerges and the resultant spectral functions give rise to lots of spurious spikes. 

Although maxEnt is currently the standard algorithm for extracting spectral information from imaginary frequency Green function, it is beset with overfitting, as illustrated above. Generally, as $\alpha \rightarrow 0$, the spectral functions get increasingly spiky, and maxEnt breaks. The traditional strategy for avoiding overfitting depends on some ingenuous strategies for selecting the optimal $\alpha$, as already noted above.  It would be more advantageous if we can eliminate the overfitting phenomenon without resort to these ingenuous yet artificial tactics. It is argued that maxEnt can be considered a regularized least square method. It is widely known that regularization can effectively reduce or even eliminate overfitting in statistics and machine learning. Inspired by the $L_1$ and $L_2$ regularization tricks in statistics, we now propose to add another regularization term to avoid overfitting in maxEnt. This regularized maxEnt method will be detailed in the next section. 

\section{Maximum entropy method with extra regularization}
\label{extra_regularization}
As noted in the previous section, classical maxEnt algorithm is vulnerable to overfitting, especially when $\alpha$ is tiny. In order to suppress this overfitting, here we propose to add a penalty term to loss function $Q$. This penalty term should be undesirable to $Q$ when the spectral function is too spiky. Based on this principle, we define the penalty term as $T = \frac{\lambda}{2}\int_{-\infty}^{\infty}\Big( A^{\prime}(\omega)\Big)^2d\omega$. Here, $\lambda > 0$ is a parameter that controls the spikiness of the resultant spectral functions. This regularization term is prohibitive when spectral functions vary rapidly and is thus analogous to the kinetic energy in classical mechanics. Similarly, the term $\chi^2$ is analogous to the potential energy in classical mechanics. Now the whole system is analogous to a harmonic oscillator since both the kinetic energy and potential energy are quadratic. With this extra regularization term, the new loss function is 
\begin{eqnarray}
Q = \frac{1}{2}\chi^2 + \frac{\lambda}{2}\int_{-\infty}^{\infty}\Big( A^{\prime}(\omega)\Big)^2d\omega - \alpha S
\end{eqnarray}
It is easy to evaluate the variation of the kinetic term with respect to $A(\omega)$, which is 
\begin{eqnarray}
\frac{\delta T}{\delta A(\omega)} &=& -\lambda \int_{-\infty}^{\infty} A^{\prime\prime}(\omega^{\prime}) \delta(\omega - \omega^{\prime}) d\omega^{\prime} \\\nonumber
&=& -\lambda A^{\prime\prime}(\omega)
\end{eqnarray}
Here, we have discarded the boundary terms due to the fact that $A(\omega = \pm\infty) = 0$.
If we discretize $\omega$, the kinetic energy term can be rewritten as 
\begin{eqnarray}
T &=& \frac{\lambda}{2} \int_{-\infty}^{\infty} \Big( A^{\prime}(\omega) \Big)^2 d\omega \\ \nonumber
&=& \frac{\lambda}{2} \sum_{\mu} \Delta\omega\Bigg( \frac{A(\omega_{\mu+1}) - A(\omega_{\mu-1})}{2\Delta\omega} \Bigg)^2
\end{eqnarray}
The derivative of this term with respect to $A(\omega_{\mu})$ is 
\begin{eqnarray}
\frac{\delta T}{\delta A(\omega_{\mu})} = -\frac{\lambda}{4\Delta\omega}\Big(A(\omega_{\mu+2}) - 2A(\omega_{\mu} ) + A(\omega_{\mu-2}) \Big)
\end{eqnarray}
The gradient of $Q$ with respect to $A(\omega_{\mu})$ is thus
\begin{eqnarray}
f(A(\omega_{\mu})) &=& \frac{\delta Q}{\delta A(\omega_{\mu})} \\\nonumber
&=& \alpha \Big(1 + \log\frac{A(\omega_{\mu})}{D(\omega_{\mu})}\Big) \Delta\omega \\\nonumber
&-& \sum_{nm} \tilde{K}_{R}(\omega_n, \omega_{\mu})\Delta\omega (\Lambda^{-1})_{nm}\Big(\tilde{G}_{R} - \tilde{K}_{R}A\Big)_m \\\nonumber
&-& \sum_{nm} \tilde{K}_{I}(\omega_n, \omega_{\mu}) \Delta\omega (\Lambda^{-1})_{nm}\Big(\tilde{G}_{I} - \tilde{K}_{I}A \Big)_{m} \\\nonumber 
&-& \frac{\lambda}{4\Delta\omega}\Big(A(\omega_{\mu+2}) - 2A(\omega_{\mu} ) + A(\omega_{\mu-2}) \Big)
\end{eqnarray}
And its Hessian matrix is 
\begin{eqnarray}
H_{\mu\nu} &=& \frac{\delta f(A(\omega_\mu))}{\delta A(\omega_\nu)} \\\nonumber
&=& \frac{\delta^2Q}{\delta A(\omega_{\mu})\delta A(\omega_{\nu})} \\\nonumber
&=& \alpha\Delta\omega\frac{\delta_{\mu\nu}}{A(\omega_\mu)} \\\nonumber
&& +  \sum_{nm}\tilde{K}_{R}(\omega_n, \omega_{\mu}) \Delta\omega(\Lambda^{-1})_{nm}\tilde{K}_{R}(\omega_m, \omega_{\nu})\Delta\omega \\\nonumber
&& + \sum_{nm}\tilde{K}_{I}(\omega_n, \omega_{\mu}) \Delta\omega(\Lambda^{-1})_{nm}\tilde{K}_{I}(\omega_m, \omega_{\nu})\Delta\omega \\\nonumber
&& - \frac{\lambda}{4\Delta\omega}\Big(\delta_{\mu + 2, \nu} - 2\delta_{\mu,\nu} + \delta_{\mu - 2, \nu} \Big)
\end{eqnarray}
This Hessian matrix is still positive definite. Thus, the new loss function remains convex and can be numerically solved using Newton's iteration method as described in the previous section. The addition of this extra regularization term requires no significant modification of the classical maxEnt program. We only need to add an extra term to the gradient and another term to the Hessian matrix to get the new maxEnt program. The new program can be found via this link: \url{https://github.com/PrimerLi/maximum-entropy-method}. 

\begin{figure}[h!]
\centerline{\includegraphics[scale = 0.35]{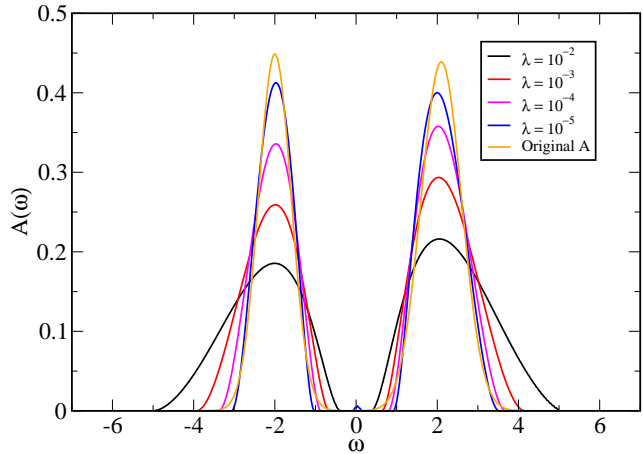}}
\caption{Comparison of the calculated spectral functions for different values of $\lambda$ with the original known spectral function. A smaller $\lambda$ gives a better fitted curve. However, setting $\lambda = 0$ will bring about overfitting problem. }
\label{A_lambda}
\end{figure}

To test the effectiveness of this program with extra regularization, we have used the same strategy as in the previous section. In Fig. [\ref{A_lambda}], we have plotted the resultant spectral functions for different values of $\lambda$. We have obtained the resultant spectral functions for $\alpha = 1.0\times 10^{-7}$. If there is no kinetic energy regularization, which is the case in the previous section, then there is severe overfitting problem. However, if we introduce the kinetic energy regularization, then the resultant spectral functions are smooth and regularized. In Fig. [\ref{A_lambda}], comparison is made for different values of $\lambda$, together with the original known spectral curve. From the figure, we can see clearly that all the spectral functions are well behaved and exhibit similar behavior. Moreover, the smaller the $\lambda$, the more close $A(\omega)$ is to the original spectral function. Setting $\lambda = 0$ will result in overfitting, and a large $\lambda$ results in a large deviation from the original spectral function. Thus, here we still need to select a hyper-parameter $\lambda$, although the final spectral result is not too sensitive to the value of $\lambda$, which renders the regularized maxEnt advantageous to the classical maxEnt algorithm.  

\section{Experimental results with real data}
In this section, we will present the program experimental results with real data. These real data are generated while we try to simulate periodic Anderson model using continuous time quantum Monte Carlo method. The Hamiltonian of periodic Anderson model is 
\begin{eqnarray}
\hat{H} &=& \hat{H}_{0} + \hat{H}_{I} 
\label{eq:Hamiltonian}\\\nonumber
\hat{H}_{0} &=& -t\sum_{\langle i, j \rangle, \sigma} (c_{i,\sigma}^{\dagger} c_{j,\sigma} + c_{j,\sigma}^{\dagger} c_{i,\sigma}) + \epsilon_{f} \sum_{i, \sigma} f_{i, \sigma}^{\dagger}f_{i, \sigma} \\\nonumber
&& + V \sum_{i,\sigma} (c_{i,\sigma}^{\dagger}f_{i,\sigma} + f_{i,\sigma}^{\dagger}c_{i,\sigma}) \\\nonumber
\hat{H}_{I} &=& U\sum_{i} n_{i,\uparrow}^{f}n_{i,\downarrow}^{f} 
\end{eqnarray}
Here, $c_{i, \sigma}^{\dagger}, c_{i, \sigma} (f_{i, \sigma}^{\dagger}, f_{i, \sigma})$ creates and destroys a $c(f)$ electron of spin $\sigma$ at lattice site $i$, respectively.  $U$ is the Hubbard repulsion between localized $f$-electrons, and $V$ characterizes the hybridization between conduction- and $f$-electrons. We have chosen the chemical potential and $\epsilon_f$ in such a way as to set the filling number of both conduction electrons and $f$ electrons to be 1, or half filling. We have solved this model using dynamical mean field approximation\cite{georges1996dynamical}, with continuous time quantum Monte Carlo as the impurity solver\cite{assaad2007diagrammatic}. We have used hyper-cubic lattice structure in the dynamical mean field approximation, and the density of states for conduction electron is Gaussian. We use the bandwidth of this Gaussian distribution as our unit of energy. Here, we present our simulation results under this unit system for parameter values $U = 4, V = 0.6$ and $T = 0.1$, where $T$ is the temperature at which the simulation is performed. We will focus attention on the spectral behavior of $f$ electrons, for which there are already well known results\cite{hewson1997kondo, pruschke2000low}. We will use these known results to benchmark our method. 

\begin{figure}[h!]
\centerline{\includegraphics[scale = 0.35]{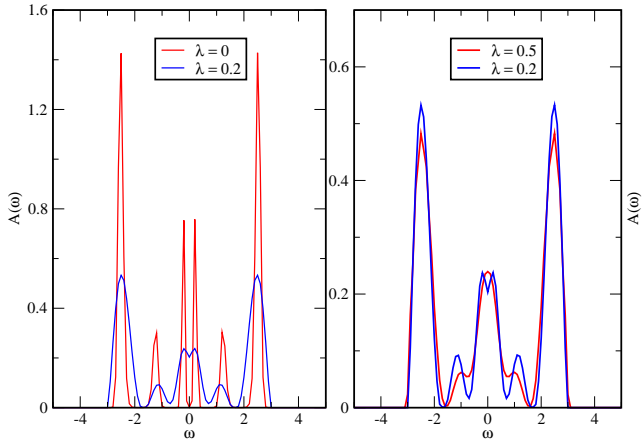}}
\caption{Spectral functions for different values of $\lambda$ with fixed $\alpha = 0.08$. See section \ref{classicalMaxEnt} and \ref{extra_regularization} for detailed meanings of $\alpha$ and $\lambda$. Note that left and right panels have different scales on their vertical axes. \textbf{Left:} Red curve for $\lambda = 0$ and blue curve for $\lambda = 0.2$. $\lambda = 0$ means no extra regularization is introduced, and thus the resultant spectral curve exhibits spurious spikes, which is an indicator of overfitting. The curve corresponding to $\lambda = 0.2$ is smoother, with fewer and less spiky peaks. \textbf{Right:} Red curve for $\lambda = 0.5$ and blue curve for $\lambda = 0.2$. Here, the red curve has a larger regularization parameter than blue curve, and is even smoother. In the red curve, spurious peaks, if they exist, are negligible. Overfitting is effectively eliminated here. }
\label{PAM_A_lambda}
\end{figure}

We have obtained $f$ electron spectral functions for different values of $\lambda$ when $\alpha$ is small enough ($\alpha \approx 0.08$). The three peak structure in the $f$ electron spectral curve is clearly visible. These three peaks correspond to the Hubbard repulsion between localized $f$ electrons ($\omega = \pm U/2$) and the Kondo peak ($\omega = 0$). The comparison of resultant spectral curve for different values of $\lambda$ are shown in Fig. \ref{PAM_A_lambda}. In the figure, we can see that when $\lambda = 0$, which means no extra regularization is introduced, the spectral curve exhibits plenty of spurious peaks, which indicates the emergence of overfitting. As $\lambda$ increases, the overfitting behavior gradually disappears. For $\lambda = 0.2$, the spurious spikes are already pretty mild, and when $\lambda = 0.5$, there are few, if any, spurious peaks. We also tested with the many other values of $\lambda$, and see that for $\lambda$ values beyond $\lambda = 0.5$, the spectral curves no longer change significantly, indicating that overfitting behavior are effectively suppressed by $\lambda$.  Moreover, the larger the $\lambda$ is, the smoother the resultant spectral curve becomes. However, a $\lambda$ that is too large may not be a good choice, since it is possible that meaningful features may simultaneously be eliminated together with the undesirable spurious spikes. 

\section{Conclusion}
In this paper, we have introduced an extra regularization term to the classical maxEnt loss function. This regularization term is meant to suppress overfitting problems in classical maxEnt. We have shown that the introduction of this regularization term can significantly enhance the final spectral functions. Another advantage is that under the new framework, we do not have to worry too much about the choice of hyper-parameters in this model since the final result is robust against hyper-parameter variation.

\begin{acknowledgments}
This material is partially based upon work supported by the National Science Foundation under the NSF EPSCoR Cooperative Agreement No. EPS-1003897 with additional support from the Louisiana Board of Regents. Computer resources for Monte Carlo simulation are provided by the Louisiana Optical Network Initiative, and by HPC@LSU computing. The author would like to thank Mr. Jie Tang and Dr. Zhengyi Le at Suning R \& D Center for the supports they cordially provide, and thank Dr. Shu-Ting Pi and Dr. Ka-Ming Tam for useful discussions. 
\end{acknowledgments}

\appendix
\section{Mathematical details of the calculation of gradient and Hessian matrix of loss function $Q$ with respect to spectral function $A(\omega)$}
In Section \ref{classicalMaxEnt}, we have defined the loss function (including the imaginary part of covariance matrix) for maxEnt, which is 
\begin{eqnarray}
\label{objective_function}
Q &=& \frac{1}{2} \chi^2 - \alpha S \\\nonumber
&=&  \frac{1}{2} \tilde{\xi}_{R}^T\Lambda^{-1}\tilde{\xi}_{R} + \frac{1}{2}\tilde{\xi}_I^T\Lambda^{-1}\tilde{\xi}_{I} + \tilde{\xi}_{I}^{T} U^T C_{I}^{-1} U\tilde{\xi}_{R} - \alpha S
\end{eqnarray}
This is a functional with $A(\omega)$ as its variable. In this section, we will ignore $C_{I}^{-1}$ which is the imaginary part of the inverse of covariance matrix. The influence of this term on the final result will be discussed in the next section. The functional dependence of $Q$ on $A(\omega)$ is hidden in $\tilde{\xi}_{R}, \tilde{\xi}_{I}$ and $S$, which are 
\begin{eqnarray}
\label{xi_definition}
&& \tilde{\xi}_{R} = U^T(\bar{G}_{R} - K_{R} A) := \tilde{G}_{R} - \tilde{K}_{R} A \\\nonumber
&& \tilde{\xi}_{I} = U^T(\bar{G}_{I} - K_{I} A) := \tilde{G}_{I} - \tilde{K}_{I} A \\\nonumber
&& S = -\int_{-\infty}^{\infty} A(\omega)\log\frac{A(\omega)}{D(\omega)}d\omega
\end{eqnarray}
Here, we can understand $\tilde{K}_R, \tilde{K}_I$ as linear operators that map spectral function in real frequency space to a vector in imaginary frequency space. The explicit expression of these two mappings are: 
\begin{eqnarray}
\label{KA_definition}
 \Big(\tilde{K}_{R} A \Big)_{n} &=& \int_{-\infty}^{\infty} \sum_{m} U_{nm} \frac{-\omega A(\omega)}{\omega_{n}^2 + \omega^2}d\omega \\\nonumber
&:=& \int_{-\infty}^{\infty}\tilde{K}_{R}(\omega_n, \omega) A(\omega)d\omega \\\nonumber
 \Big( \tilde{K}_{I} A \Big)_{n} &=&  \int_{-\infty}^{\infty} \sum_{m} U_{nm} \frac{-\omega_{n} A(\omega)}{\omega_{n}^2 + \omega^2}d\omega \\\nonumber
 &:=& \int_{-\infty}^{\infty}\tilde{K}_{I}(\omega_n, \omega) A(\omega)d\omega
\end{eqnarray}

Variation of the first term in $Q$ with respect to $A(\omega)$ yields 

\begin{widetext}
\begin{eqnarray}
&& \frac{\delta}{\delta A(\omega)} \Bigg((\tilde{G}_{R} - \tilde{K}_{R} A )^T\Lambda^{-1}(\tilde{G}_{R} - \tilde{K}_{R} A)\Bigg)\\\nonumber
&=& \frac{\delta}{\delta A(\omega)}\sum_{mn}\Bigg(\Big(\bar{G}_{R}(\omega_m) - \int_{-\infty}^{\infty}\tilde{K}_{R}(\omega_m, \omega)A(\omega)d\omega \Big) \Lambda^{-1}_{mn} \Big( \bar{G}_{R}(\omega_n) - \int_{-\infty}^{\infty}\tilde{K}_{R}(\omega_n, \omega)A(\omega)d\omega\Big) \Bigg)\\\nonumber
  &=& -\sum_{mn}\tilde{K}_{R}(\omega_m, \omega)\Lambda^{-1}_{mn} \Big( \bar{G}_{R}(\omega_n) - \int_{-\infty}^{\infty}\tilde{K}_{R}(\omega_n, \omega)A(\omega)d\omega\Big) \\\nonumber
  && - \sum_{mn}\Big( \bar{G}_{R}(\omega_m) - \int_{-\infty}^{\infty}\tilde{K}_{R}(\omega_m, \omega)A(\omega)d\omega\Big) \Lambda^{-1}_{mn} \tilde{K}_{R}(\omega_m, \omega)\\\nonumber
   &=& -2\sum_{mn} \tilde{K}_{R}(\omega_m, \omega)\Lambda^{-1}_{mn} \Big( \bar{G}_{R}(\omega_n) - \int_{-\infty}^{\infty}\tilde{K}_{R}(\omega_n, \omega)A(\omega)d\omega\Big)
\end{eqnarray}
The total variation of $\frac{1}{2}\chi^2$ with respect to $A(\omega)$ is 
\begin{eqnarray}
&& \frac{\delta}{\delta A(\omega)}\frac{1}{2}\chi^2 \\\nonumber
& =&  -\sum_{mn}\tilde{K}_{R}(\omega_m, \omega)\Lambda^{-1}_{mn} \Big( \bar{G}_{R}(\omega_n) - \int_{-\infty}^{\infty}\tilde{K}_{R}(\omega_n, \omega)A(\omega)d\omega\Big) \\\nonumber
&& - \sum_{mn} \tilde{K}_{I}(\omega_m, \omega)\Lambda^{-1}_{mn} \Big( \bar{G}_{I}(\omega_n) - \int_{-\infty}^{\infty}\tilde{K}_{I}(\omega_n, \omega)A(\omega)d\omega\Big)
\end{eqnarray}
\end{widetext}

In order to find the spectral function that could minimize $Q$, we should solve this equation: 
\begin{eqnarray}
\frac{\delta Q}{\delta A(\omega)} = 0
\end{eqnarray}
If we also introduce the kinetic energy regularization term, that is, redefine loss function as 
\begin{eqnarray}
Q = \frac{1}{2}\chi^2 + \frac{\lambda}{2}\int_{-\infty}^{\infty}\Big( A^{\prime}(\omega)\Big)^2d\omega - \alpha S, 
\end{eqnarray}
then the gradient of $Q$ with respect to $A(\omega)$ becomes
\begin{widetext}
\begin{eqnarray}
\frac{\delta Q}{\delta A(\omega)} &=& \alpha \Big(1 + \log\frac{A(\omega)}{D(\omega)}\Big)  \\\nonumber
&& -\sum_{mn}\tilde{K}_{R}(\omega_m, \omega)\Lambda^{-1}_{mn} \Big( \bar{G}_{R}(\omega_n) - \int_{-\infty}^{\infty}\tilde{K}_{R}(\omega_n, \omega)A(\omega)d\omega\Big) \\\nonumber
&& - \sum_{mn} \tilde{K}_{I}(\omega_m, \omega)\Lambda^{-1}_{mn} \Big( \bar{G}_{I}(\omega_n) - \int_{-\infty}^{\infty}\tilde{K}_{I}(\omega_n, \omega)A(\omega)d\omega\Big) \\\nonumber 
&& -\lambda A^{\prime\prime}(\omega) \\\nonumber
&=& 0
\end{eqnarray}
\end{widetext}
This is a differential-integral equation. Solution of this equation requires Newton's iteration method, which further requires knowledge of Hessian matrix of $Q$, which is
\begin{eqnarray}
H_{\omega, \omega^{\prime}} &=& \frac{\delta^2 Q}{\delta A(\omega)\delta A(\omega^{\prime})} \\\nonumber
&=& \alpha\frac{\delta(\omega - \omega^{\prime})}{A(\omega)} \\\nonumber
&&  + \sum_{mn} \tilde{K}_{R}(\omega_m, \omega)\Lambda^{-1}_{mn}  \tilde{K}_{R}(\omega_n, \omega^{\prime}) \\\nonumber
&&  + \sum_{mn} \tilde{K}_{R}(\omega_m, \omega)\Lambda^{-1}_{mn}  \tilde{K}_{R}(\omega_n, \omega^{\prime}) 
\end{eqnarray}
Here we have ignored the term $-\lambda A^{\prime\prime}(\omega)$, since the variation of this term with respect to $A(\omega)$ is intuitively meaningful only when $\omega$ is discretized, and the Hessian matrix for discretized $\omega$ is already given in Section \ref{extra_regularization}. 

Up to now, we have derived the gradient and Hessian matrix of $Q$ in continuous real frequency space. The derivations of these two formulae in discrete real frequency space are straightforward and will be omitted. 

\section{Discussion of the influence of the imaginary part of covariance matrix upon maxEnt method}
The imaginary part of covariance matrix enters into the loss function in the form $\tilde{\xi}_{I}^{T} U^T C_{I}^{-1} U\tilde{\xi}_{R}$, where $C_{I}^{-1}$ is a real and anti-symmetric matrix. The definitions of $\tilde{\xi}_{R}, \tilde{\xi}_{i}, \tilde{K}_{R} A, \tilde{K}_{I} A$ are given in Equ.  [\ref{xi_definition}] and [\ref{KA_definition}]. The variation of this term with respect to spectral function $A(\omega)$ can also be obtained using the method detailed in the previous section. However, note that here the matrix $U^T C_{I}^{-1} U $ is not necessarily diagonal and we need to deal with unwieldy non-diagonal matrices. We make the claim that the influence of this term upon maxEnt is negligible when the correlation between Green functions from different bins is weak, because in the limiting case where there is no correlation between different bins, the covariance matrix is diagonal, and the diagonal elements are just the variances of each bin-averaged $G(i\omega_n)$. We know that the covariance matrix is Hermitian, and thus a diagonal covariance must be real. As a result, when there is no correlation between bins, the imaginary part of covariance matrix is identically zero. Consequently, we can safely deduce that when the correlation across different bins is weak, the imaginary part of covariance would be negligible. This deduction justifies our ignorance of the imaginary part of covariance during the program implementation of maxEnt, as long as we are careful to make sure that the Monte Carlo bin size is large enough so that the correlation across different bins is indeed negligible. According to the Monte Carlo simulation results for symmetric periodic Anderson model, the Frobenius norm of the real part of covariance matrix is several orders of magnitude larger than that of the imaginary part, thus substantiating our assumption. 

\bibliography{ref}
\bibliographystyle{apsrev}

\end{document}